# Observing NY Vir and the quest for circumbinary planets

(David Pulley, George Faillace, Derek Smith, Americo Watkins)

For amateur astronomers the eclipsing binary NY Vir is an ideal object to observe. Relatively bright at 13$^{th}$ magnitude and positioned just below the celestial equator, it makes for comparatively easy viewing from the UK through spring. Its short period of just less than 2.5 hours makes it possible to record a full light curve in a single observing session. NY Vir is one of eight sdB eclipsing binaries we have been monitoring over the past four years to track variations in period which, amongst other things, may be indicative of circumbinary object(s).

NY Vir was identified in 1998 by South African professional David Kilkenny as a hot subdwarf (sdB) pulsating star with a late dwarf M5 type companion. He assigned a linear ephemeris to the eclipse times but noted this was probably an approximation. By 2011 Kilkenny added a further 12 times of minima noting that the binary period appeared to be changing rapidly. To explain this deviation from linear he added a quadratic term to the ephemeris. There are a number of possible causes for the predicted times of minima differing from the linear calculated values including (i) angular momentum loss through magnetic breaking or gravitational wave generation (ii) star spots (e.g. Applegate effect) (iii) light travel time effects caused by the binary pair rotating around the system's barycentre in the presence of one or more circumbinary objects (much like our Sun wobbles around our Solar System barycentre). This causes the system eclipses to sometimes occur nearer to the Earth than at other times causing the eclipses to appear to occur earlier or later than expected.

With more data being added by other researchers it was the Chinese astronomer S-B Qian, who suggested in 2012, that this system showed a cyclical change with a 7.9 year period possibly implying the presence of a circumbinary planet of mass ~ 2.3M$_J$. Qian also put forward the possibility that the quadratic nature of the ephemeris could be better explained by a long period circumbinary second planet. These predictions were made by studying the difference in time between when an eclipse occurred and when the linear ephemeris predicted it. If these time differences are plotted against the eclipse number (epoch), the resulting pattern can be analysed for periodic changes and so predict the possible presence of circumbinary bodies.

In 2014 Jae Woo Lee's team added further times of minima and calculated parameters for two circumbinary planet(s) orbiting the binary system. Their first planet aligned well with Qian's proposal with a predicted mass of 2.78M$_J$ and period of 8.18 years. They were unsure of a second planet but suggested a mass of 4.49M$_J$ and 27 year period. Their analysis and data points used are shown as the red squares in observed minus calculated (O – C) plot in Fig. 1. We have also included the six data points from David Kilkenny's recent 2014 paper. The red line is

Lee's predicted best fit to residual data based upon their two planet hypothesis combing the two sinusoidal orbits with periods given above.

From our observations listed in Table 1, we have added a further nine times of minima between April 2015 and May 2016 which are the blue diamonds to the extreme right in Fig. 1. Clearly our new data does not fit with the predictions of Lee and unlikely to fit those of Qian. Furthermore the quadratic hypothesis first proposed by Kilkenny is also unlikely to fit our new data. We did question the validity of our data, particularly our timing accuracy, time conversion from $JD_{UTC}$ to $BJD_{TDM}$ and minima calculation. Our data points were collected from four independent telescopes suggesting timing issues to be unlikely. Our time conversion employed the widely used routines from Ohio State University and times of minima calculated from the established Kwee and van Woerden routine encoded in Peranso. Could we have made a mistake?

We have now identified four new datasets on the AAVSO website which give added confidence in our observations. The AAVSO database provided the raw data for four primary minima at E = 27769, 27770, 39449 and 42934. Our analysis of the AAVSO data found their first two points falling within Lee's time frame and fitted well to their predicted curve and their latter two data points falling within our time frame and fitting well with our data. This is more clearly shown in Fig.2.

With this confirmation we conclude that the one planet scenario of Qian and the two planet scenario of Lee both probably need revisiting; similarly the quadratic term proposed for the ephemeris is likely to be based upon a wrong assumption. So what has caused these real period variations? That remains an open question and more data is needed but we will be publishing a more detailed report on NY Vir and seven other sdB binaries.

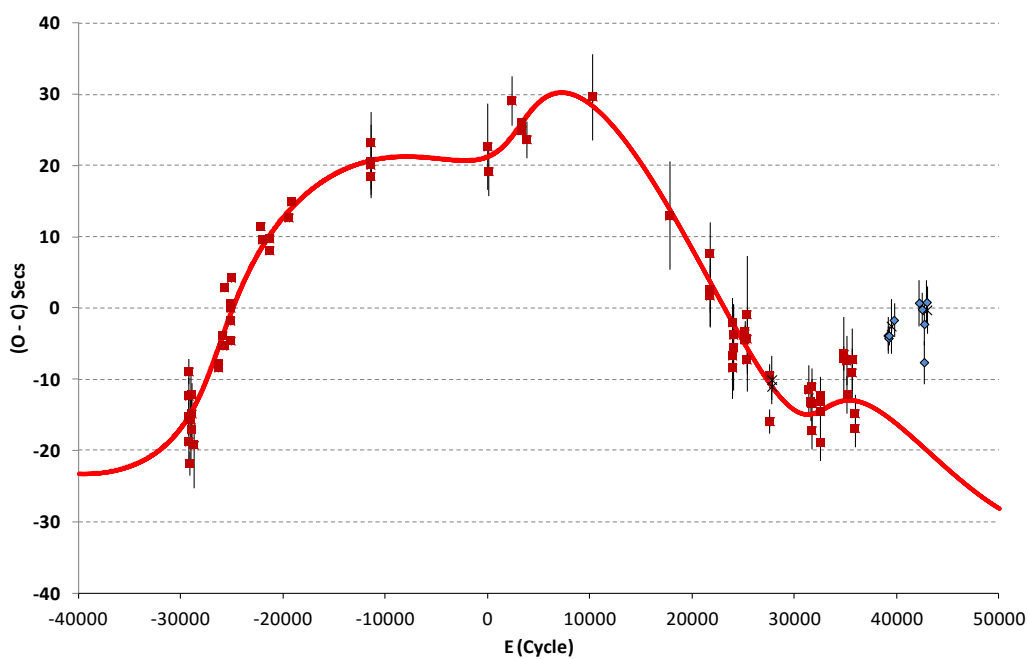

Fig. 1 Plot of residuals for NY Vir. Red squares are data used by Lee and Kilkenny's 2014 data; Blue diamonds are our new data; Red line is Lee's prediction for two circumbinary planets.

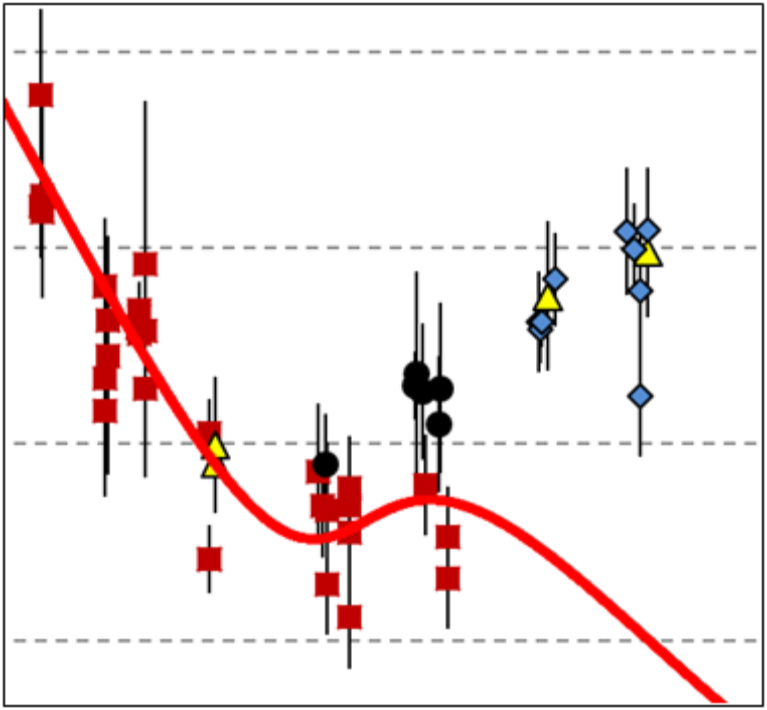

Fig. 2: Enlarged view between 25,000< E < 45,000 displaying the last few data sets used by Lee (red squares), our new data (blue diamonds) and the four data sets extracted from the AASVO database (yellow triangles). Kilkenny's recent data is also shown (black circles) starting to deviate above the predicted curve

| JD (+2400000) | BJD (+2400000) | Error +/- | E (Cycle) | Filter | Observer | Telescope |
|---|---|---|---|---|---|---|
| 57124.867573 | 57124.874069 | 0.000034 | 39107 | Clear | DP | 0.61m SSO, Markleeville, Ca, USA |
| 57132.948864 | 57132.955342 | 0.000021 | 39187 | Clear | DP | 2m Faulkes, Hawaii [1] |
| 57136.484469 | 57136.490904 | 0.000015 | 39222 | Clear | GF | 0.32m T18 iTelescope, Nerpio, Spain |
| 57185.782811 | 57185.786722 | 0.000028 | 39710 | Clear | DP | 0.61m SSO, Markleeville, Ca, USA |
| 57434.989263 | 57434.993139 | 0.000035 | 42177 | Clear | DP | 0.61m SSO, Markleeville, Ca, USA |
| 57463.978897 | 57463.984711 | 0.000027 | 42464 | Clear | DP | 0.61m SSO, Markleeville, Ca, USA |
| 57483.474275 | 57483.480707 | 0.000036 | 42657 | Sloan r' | AW | 0.36m Astrognosis Obs., Essex |
| 57485.494629 | 57485.501089 | 0.000034 | 42677 | Clear | AW | 0.36m Astrognosis Obs., Essex |
| 57509.839708 | 57509.845972 | 0.000037 | 42918 | Clear | DP | 0.61m SSO, Markleeville, Ca, USA |

Table 1. Our new times of minima observed between April 2015 and May 2016. Telescopes: SSO ~ Sierra Stars Observatory: Astrognosis ~AW's observatory. Observers initials are those of the authors.

(1) This work makes use of observations from the LCOGT network of telescopes

## References

Kilkenny et al., 1998, MNRAS, 296, 329

Kilkenny, 2011, MNRAS, 412, 489

Kilkenny, 2014, MNRAS, 445, 4247

Kwee & van Woerden, 1956, Bull. Astron. Inst., Netherlands, 12, 327

Lee et al., 2014, MNRAS, 445, 2331

Qian et al., 2012, ApJ, 745, L23